\documentclass[12pt,preprint]{aastex}
\shortauthors{Ryu \& Lee}
\begin{document}
\title{Constraining the Neutrino Mass with the Drifting Coefficient of the Field Cluster Mass Function}
\author{Suho Ryu and Jounghun Lee}
\affil{Astronomy program, Department of Physics and Astronomy,
Seoul National University, Seoul  08826, Republic of Korea \\
\email{shryu@astro.snu.ac.kr, jounghun@astro.snu.ac.kr}}
\begin{abstract}
A new diagnostics to break the degeneracy between the total neutrino mass ($M_{\nu}$) and the primordial power spectrum amplitude ($\sigma_{8}$) 
by using the drifting coefficient of the field cluster mass function is presented. Analyzing the data from the Cosmological Massive Neutrino Simulations, 
we first determine the numerical mass functions of the field clusters at various redshifts. Then, we compare the numerical 
results with the analytical model characterized by a single parameter called the drifting coefficient which measures the drifts of the collapse density
threshold,   $\delta_{c}$, from the Einstein-de Sitter spherical value, $\delta_{sc}$, at a given mass scale.  It is found that the analytic model for the field 
cluster mass function is found to work excellently even in the presence of massive neutrinos and that its drifting coefficient evolves differently in  
the cosmologies with different values of $M_{\nu}$.  At low redshifts ($z\lesssim 0.3$) the more massive neutrinos drift $\delta_{c}$ further from $\delta_{sc}$, 
while the opposite trend is found at higher redshifts ($z\gtrsim0.3$). Speculating that this distinct redshift-dependent effect of massive neutrinos on the drifting 
coefficient of the field cluster mass function might help break the $\sigma_{8}$-$M_{\nu}$ degeneracy, we also show that the sensitivity 
of this new diagnostics to $M_{\nu}$ is high enough to discriminate the case of $M_{\nu}=0.1\,{\rm eV}$ from that of massless neutrinos. 
\end{abstract}
\section{Introduction}\label{sec:intro}

The galaxy clusters are often divided into two categories, the wall and the field clusters to which the members of the superclusters and the rest 
correspond, respectively. 
Presuming that the field clusters are more isolated and thus less susceptible to the disturbing effects of the surrounding cosmic web than their wall 
counterparts, \citet{lee12} claimed that the field clusters should provide a more sensitive indicator of the background cosmology, 
developing an analytic model for the field clusters in the generalized excursion set framework \citep{MR10a,MR10b,CA11a,CA11b}.
The analytic model of \citet{lee12} was indeed a practical and theoretical success.  On the practical side, its success was demonstrated by 
the excellent agreements with the N-body results obtained assuming a base cosmology, where the late-time cosmic acceleration is caused by the 
cosmological constant ($\Lambda$) and the structure formation is predominantly driven by the gravity of the cold dark matter (CDM).

On the theoretical side, its success resonates with the fact that it is a physical model having only one deterministic parameter.  
Unlike the conventionally used empirical formulae with physically meaningless multiple stochastic parameters for the cluster mass function 
\citep[e.g.,][]{ST99,war-etal06,tin-etal08},  
the analytic model of \citet{lee12} for the field cluster mass function imparts physical substance to its {\it single} parameter. 
Although it is inevitable to resort to the numerical experiments for the determination of its exact value, this single parameter dubbed 
{\it drifting coefficient} \citep{CA11a} is a physical measure of how far the density threshold for the realistic non-spherical collapse, $\delta_{c}$, 
drifts from that for the idealistic spherical collapse, $\delta_{sc}$, at a given mass scale. 

Moreover, the recent work of \citet{RL20} discovered that the drifting coefficient of the field cluster mass function in fact carries significance 
beyond a physical parameter quantifying the deviation of $\delta_{c}$ from $\delta_{sc}$. Confirming that the validity of the analytic model 
of \citet{lee12} for the field cluster mass functions is robust even against the variations of the key cosmological parameters including the dark 
energy (DE) equation of state, \citet{RL20} found that the drifting coefficient evolves differently even among those degenerate dynamical DE 
cosmologies which yield almost the same linear growth factors or the same  cluster mass functions.  In the light of the results of \citet{RL20}, 
we speculate that the evolution of the drifting coefficient might be also useful to constrain the total mass of neutrino species, 
$M_{\nu}\equiv \sum m_{\nu}$, one of the utmost missions entrusted to the cosmological physics \citep{LP12}.

The latest Planck analysis of the Cosmic Microwave Background (CMB) temperature power spectra combined with the priors from the weak 
gravitational lensing (WL) and Baryonic Acoustic Oscillations (BAO) concluded $M_{\nu}\le 0.12\,{\rm eV}$ \citep{planck18}, 
assuming the base flat $\Lambda$CDM cosmology \citep[see also][]{vag-etal17}.  
A higher value of $M_{\nu}$ above $0.12\,{\rm eV}$, however, can still be accommodated by the Planck data, if the 
assumption about the background cosmology is released \citep[see][and references therein]{CC18,CH19} or if different priors are used to complement 
the CMB probe \citep[e.g.,][]{giu-etal16}. For the past decade, the cluster mass function has been prevalently promoted as an useful complementary 
probe of $M_{\nu}$ \citep[e.g.,][]{mar-etal11,IT12,cos-etal13,vil-etal13,cas-etal14,bis-etal19,hag-etal19}.  
Although the cluster mass function is only indirectly linked to $M_{\nu}$ through its dependence on the linear density power spectrum, 
it has a practical advantage as a probe of $M_{\nu}$, being more readily observable than the linear density power spectrum, the 
measurements of which are often plagued by the systematics stemmed from the existence of nonlinear galaxy bias 
\citep[][and references therein]{giu-etal18}.

Due to the inherent non-sphericity and stochastic aspect of the cluster formation process that defies purely analytic modeling from the first 
principle, a theoretical prediction for the cluster abundance and its dependence on $M_{\nu}$ was conventionally made in the {\it empirically} 
modified excursion set formalism \citep[e.g.,][]{cos-etal13,vil-etal13,bis-etal19}. 
While a link between the cluster abundance and $M_{\nu}$ through the linear power spectrum is provided by the excursion set theory, 
the required accuracy and precision was achieved by the empirical modification of the theory, i.e., deteriorating of a physical model into a 
fitting formula with multiple free parameters \citep{war-etal06,tin-etal08}.
Lack of a physical model for the cluster abundance undermines its power as a probe of $M_{\nu}$. To make matters worse, the notorious 
$\sigma_{8}$-$M_{\nu}$ degeneracy of the initial density power spectrum translates into the relative low sensitivity of the cluster mass function 
to $M_{\nu}$. 

Given the aforementioned difficulties in constraining $M_{\nu}$ with the cluster abundance, what may be desirable to have is a new probe, 
well described by a physical model, free from the $\sigma_{8}$-$M_{\nu}$ degeneracy, and highly sensitive to the variation of $M_{\nu}$.
Our goal here is to prove that the drifting coefficient of the field cluster mass function fulfills this expectation.  Section \ref{sec:review} will be 
consigned to a brief review of the works of \citet{lee12} and \citet{RL20}. Section \ref{sec:analysis} will present a procedure through which the 
power and efficacy of the drifting coefficient as a new probe of $M_{\nu}$ is numerically appraised. Section \ref{sec:con} will be devoted 
to discussing a physical implication of the final results and a prospect for constraining $M_{\nu}$ with this new probe in practice, as well.

\section{Review of the Analytic Model}\label{sec:review}

The differential mass function of the field clusters, $dN_{\rm I}/d\ln M$, gives their number densities in each  logarithmic interval of 
$[\ln M, \ln M + d\ln M]$.  Suggesting for the first time that the mass function of the field clusters should be more sensitive to the 
background cosmology than that of all clusters, \citet{lee12} modified the generalized excursion set mass function theory 
\citep{MR10a,MR10b,CA11a, CA11b} to derive the following single parameter model for $dN_{\rm I}/d\ln M$, which will be adopted for our analysis.
\begin{eqnarray}
\label{eqn:dndlnm}
\frac{d\,N_{I}(M, z)}{d\,{\rm ln}\,M} &=&  \frac{\bar{\rho}}{M}\Bigg{\vert}\frac{d\,{\rm ln}\,\sigma^{-1}}{d\,{\rm ln}\,M}\Bigg{\vert}
\left[f^{(0)}(\sigma ; \beta) + f^{(1)}_{\beta=0}(\sigma) + f^{(1)}_{\beta}(\sigma ; \beta) + f^{(1)}_{\beta^2}(\sigma ; \beta)\right]\, ,\\
f^{(0)}(\sigma ; \beta) &=& \frac{\delta_{sc}}{\sigma} \sqrt{\frac{2}{\pi}}\,
e^{-\frac{(\delta_{sc}+\beta \sigma^2)^2}{2\sigma^2}}\, ,\\
\label{eqn:f1b0}
f^{(1)}_{\beta=0}(\sigma) &=&-\kappa\frac{\delta_{sc}}{\sigma}
\sqrt{\frac{2}{\pi}}\left[e^{-\frac{\delta_{sc}^2}{2\sigma^2}}
-\frac{1}{2}\Gamma\left(0,\frac{\delta_{sc}^2}{2 \sigma^2}\right)\right]\, , \\
\label{eqn:f1b1}
f^{(1)}_{\beta}(\sigma ; \beta) &=&
-\beta\,\delta_{sc}\left[f^{(1)}_{\beta=0}(\sigma)+\kappa\,
\textrm{erfc}\left(\frac{\delta_{sc}}{\sqrt{2}\sigma}\right)\right]\, , \\
\label{eqn:f1b2}
f^{(1)}_{\beta^2}(\sigma ; \beta) &=&\beta^{2}\delta^{2}_{sc}\kappa
\biggl\{\textrm{erfc}\left(\frac{\delta_{sc}}{\sqrt{2}\sigma}\right)+\frac{\sigma}{\sqrt{2\pi}\delta_{sc}}\left[e^{-\frac{\delta_{sc}^2}
{2\sigma^2}}\left(\frac{1}{2}-\frac{\delta_{sc}^2}{\sigma^2}\right)+\frac{3}{4}\frac{\delta_{sc}^2}
{\sigma^2}\Gamma\left(0,\frac{\delta_{sc}^2}{2 \sigma^2}\right)\right]\biggr\}\, ,
\end{eqnarray}
where $\kappa = 0.475$ and $\beta$ is the drifting coefficient that quantifies how much the non-sphericity of the gravitational 
collapse drifts the density threshold $\delta_{c}$ away from the Einstein de Sitter spherical collapse threshold of $\delta_{sc}=1.686$ 
\citep{GG72, pee93} at a given mass scale.  Since the non-spherical gravitational collapse process is too complicated for $\delta_{c}$ 
to be theoretically predicted from the first principle \citep{BM96}, $\beta$ has to be treated as a free adjustable parameter, as in the generalized 
excursion set formalism \citep{CA11a,CA11b}. Nevertheless, as \citet{lee12} and \citet{RL20} explained, the density threshold $\delta_{c}$ 
(or equivalently $\beta$) is deterministic for the field clusters, while it is stochastic fo their wall counterparts \citep{rob-etal09,MR10a,MR10b}, 
which allows the field cluster mass function to have only one free parameter in the generalized excursion set formalism.

This analytical single parameter model, Equations (\ref{eqn:dndlnm})-(\ref{eqn:f1b2}), connects $dN_{\rm I}/d\ln M$ to the initial conditions of the 
universe through two different routes. The rms density fluctuation of the initial density field, $\sigma(M,z)$, expressed in terms of the linear density 
power spectrum, $P(k,z)$, is the usual route envisaged by the original excursion set mass function theory \citep{PS74,bon-etal91}. While, 
the drifting coefficient, $\beta(z)$, is another independent route induced by the cosmology dependence of $\delta_{c}$ \citep{RL20}. 
To effectively describe different behaviors of $\beta(z)$ among different DE cosmologies, the following fitting formula was proposed by 
\citet{RL20}, 
\begin{equation}
\beta(z) = \beta_{A}\ {\sinh}^{-1} \left[\frac{1}{q_z}(z-z_c)\right],
\label{eqn:beta_fit} 
\end{equation}
with three fitting parameters, $\beta_{A},\ q_{z}$ and $z_{c}$.  

As mentioned in Section \ref{sec:intro}, \citet{RL20} tested this fitting formula for $\beta(z)$ 
as well as the above analytical single parameter model for $dN_{\rm I}/d\ln M$ against the large N-body simulations for various DE cosmologies 
including the $\Lambda$CDM and confirmed that it is quite valid regardless of the DE equation of states. Moreover, it was also shown by 
\citet{RL20} that $\beta(z)$, via Equation (\ref{eqn:beta_fit}), allows us to distinguish even among those degenerate DE cosmologies which produce 
almost the same linear density power spectra and cluster mass functions. 
In Section \ref{sec:analysis}, we are going to numerically test if Equations (\ref{eqn:dndlnm})-(\ref{eqn:f1b2}) are also valid for the $\nu\Lambda$CDM 
cosmology and to examine its power as a complementary probe of $M_{\nu}$. 

Before proceeding further, it is worth mentioning that there is actually another route other than $P(k,z)$ and $\delta_{c}$ that connects 
$dN_{\rm I}/d\ln M$ to the background cosmology. This third route is nothing but the spherical collapse density threshold, $\delta_{sc}$, for which 
even purely theoretical predictions from the first principles can be made thanks to the spherical symmetry.  
The cosmology dependence of $\delta_{sc}$, however, was found too weak to stand out over those of $P(k,z)$ and $\delta_{c}$ 
\citep[e.g., see][]{eke-etal96,pac-etal10}, which is why the connection between $dN_{\rm I}/d\ln M$ and the initial conditions can be almost entirely 
attributed to $P(k,z)$ and $\delta_{c}$.  The same argument applies to the $M_{\nu}$-dependence of $\delta_{sc}$, which was already shown to 
be not so strong as that of $P(k,z)$ \citep[e.g.,][]{lov14}.  
Throughout this Letter as in \citet{RL20}, we set the spherical density threshold $\delta_{sc}$ at the fixed Einstein-de Sitter value, 
$1.686$ \citep[see also][]{MR10a,MR10b,lee12}.  

\section{The Effect of Massive Neutrinos on $\beta(z)$}\label{sec:analysis}

We make an extensive use of the publicly available data from the Cosmological Massive Neutrinos Simulations (\texttt{MassiveNuS}) run by 
\citet{massivenu} on a periodic box of comoving volume $512^{3}\,h^{-3}\,$Mpc$^{3}$, containing $1024^{3}$ particles, each of which is as massive 
as $10^{10}\,h^{-1}\,M_{\odot}$. The \texttt{MassiveNuS} was recurringly performed for one $\Lambda$CDM cosmology with massless neutrinos and 
for $100$ different $\nu\Lambda$CDM cosmologies with massive neutrinos, whose initial conditions were described by the six key cosmological 
parameters as well as $M_{\nu}$. 
For the study of the {\it sole} effect of the massive neutrinos on $dN_{\rm I}/d\ln M$ and $\beta(z)$, we consider only those cosmologies which have 
identical initial conditions other than $M_{\nu}$ with one another. Among the $101$ cosmologies are found only $3$ to 
meet this selection criterion, which have the same matter density parameter, $\Omega_{m}=0.3$, and same amplitude of the primordial density 
power spectrum, $A_{s}=2.1\times 10^{9}$, but different total neutrino mass, $M_{\nu}=0.0,\ 0.1$ and $0.6$ eV, respectively. 

The \texttt{MassiveNuS} engaged the Rockstar algorithm \citep{rockstar} to find the DM halos at various redshifts and recorded such key properties of each 
Rockstar halo as its virial mass ($M$), virial radius, comoving position vector, peculiar velocity vector and so forth. 
From the catalog of the Rockstar halos resolved at each redshift for each of the three cosmologies, we first exclude the subhalos embedded in larger parent 
halos and then set the cutoff mass at  $3\times 10^{13}\,h^{-1}\,M_{\odot}$ to sort out the distinct cluster halos. Following the same procedure arranged 
in \citet{RL20}, we apply the friends-of-friends (FoF) algorithm with the linkage length parameter of $l_{\rm c}=0.33$ to the distinct cluster halos for the 
identification of the superclusters composed of two or more members.
Eliminating the wall clusters belonging to the identified superclusters, we end up having a sample of the {\it distinct field cluster halos} with 
$M\ge 3\times 10^{13}\,h^{-1}\,M_{\odot}$. Then, we reckon the field clusters at each logarithmic mass bin to numerically determine $dN_{I}/d\ln M$ 
to which the analytical single parameter model,  Equations (\ref{eqn:dndlnm})-(\ref{eqn:f1b2}), is fitted by adjusting the value of $\beta$.  

In the procedure of evaluating the analytic mass functions of the field clusters, the CAMB code \citep{camb} is exclusively used for $P(k,z)$, while the 
standard $\chi^{2}$-statistics is employed for the best-fit value of $\beta$. Note that since the linear growth factor, $D(z)$, acquires a scale dependence 
in the presence of massive neutrinos, the rms density fluctuation $\sigma(M,z)$, is no longer equal to $D(z)\sigma(M,z=0)$.  
Instead, we calculate it as $\sigma(M,z)=\left[(2\pi^{2})^{-1}\int dk\,k^{2}P(k,z)W^{2}_{\rm th}(k,M) \right]^{1/2}$ where $W_{\rm th}$ is the spherical 
top-hat filter on the mass scale of $M$.

\begin{figure}
\begin{center}
\includegraphics[scale=0.7]{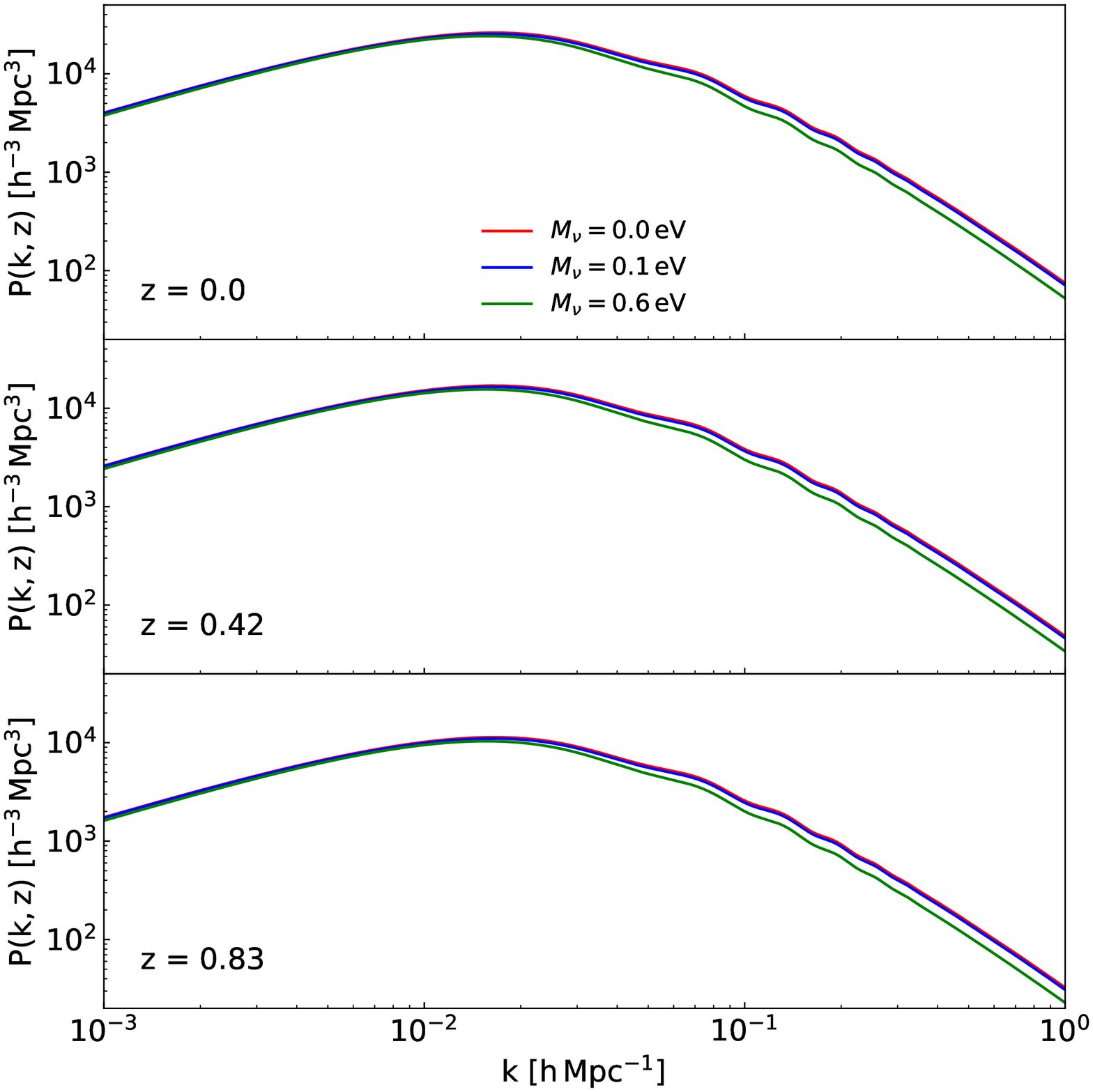}
\caption{Linear density power spectra for three different values of total neutrino mass 
($M_{\nu}=0.0,\ 0.1,\ 0.6$ eV) at three different redshifts ($z=0.0,\ 0.42,\ 0.83$), computed by the CAMB code \citep{camb}.}
\label{fig:pk}
\end{center}
\end{figure}

Figure \ref{fig:pk} plots the linear density power spectra, $P(k,z)$, for the three different cases of $M_{\nu}$ at three different redshifts, computed 
by the CAMB code. As expected, the more massive neutrinos suppress more severely the linear density powers on the small scales 
($k> 0.02\,h\,{\rm Mpc}^{-1}$). 
Note the small differences in $P(k,z)$ between the cases of $M_{\nu}=0.0$ eV and $M_{\nu}=0.1$ eV at all of the three redshifts. Given that the large 
uncertainties in the high-mass tails of the cluster mass functions caused by poor-number statistics and cosmic variance are likely to exceed this small 
differences in $P(k,z)$, the cluster mass functions would be unable to discriminate the two $\nu\Lambda$CDM cosmologies from each other.

\begin{figure}
\begin{center}
\includegraphics[scale=0.7]{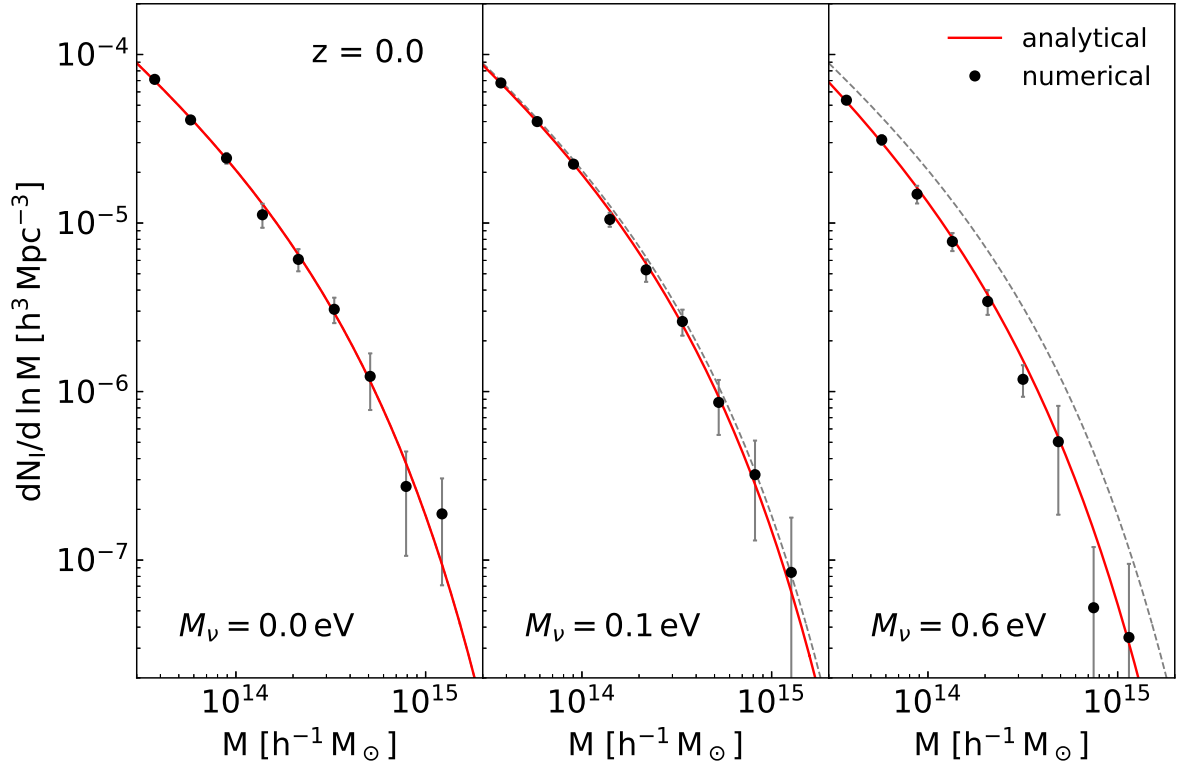}
\caption{Analytic mass functions of the field clusters (red solid lines) over-plotted with the numerical 
results from the \texttt{MassiveNuS} for the three different cases of $M_{\nu}$ at $z=0$. The dotted 
lines in the middle and right panels conform to the red solid line in the left panel.}
\label{fig:mf_z0}
\end{center}
\end{figure}

\begin{figure}
\begin{center}
\includegraphics[scale=0.7]{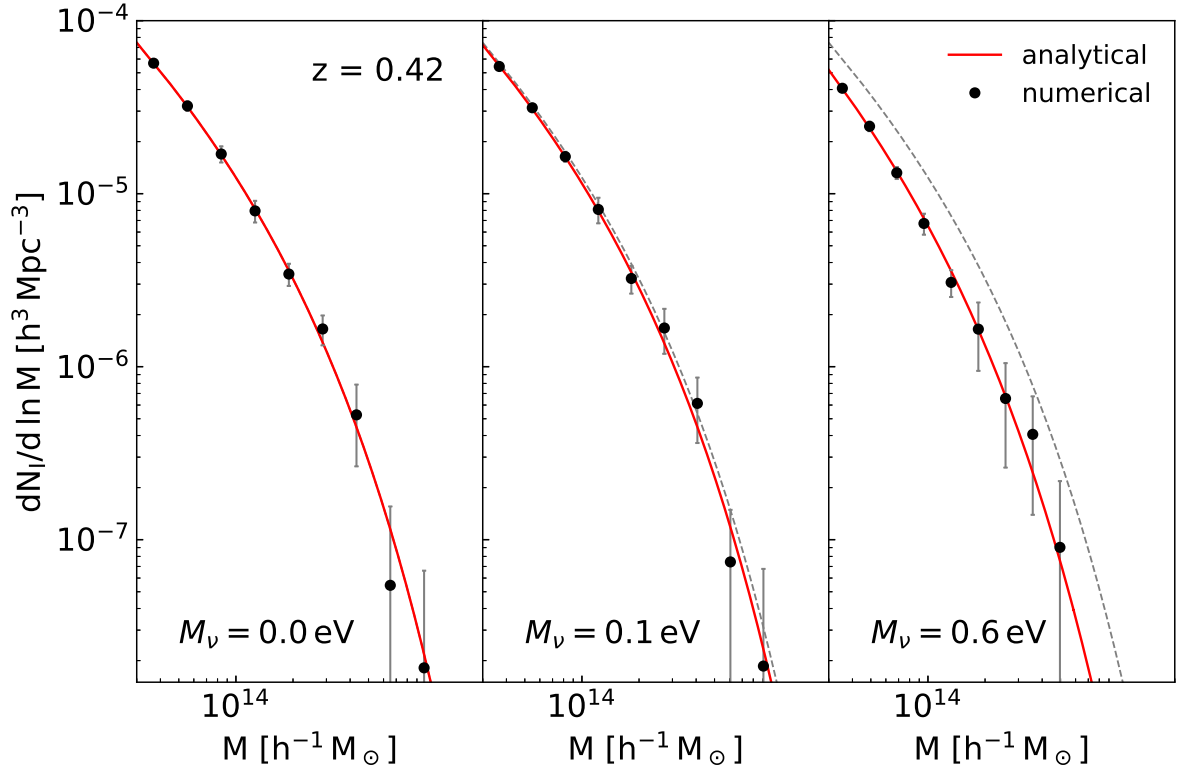}
\caption{Same as Figure \ref{fig:mf_z0} but for at $z=0.42$.}
\label{fig:mf_z0.4}
\end{center}
\end{figure}

\begin{figure}
\begin{center}
\includegraphics[scale=0.7]{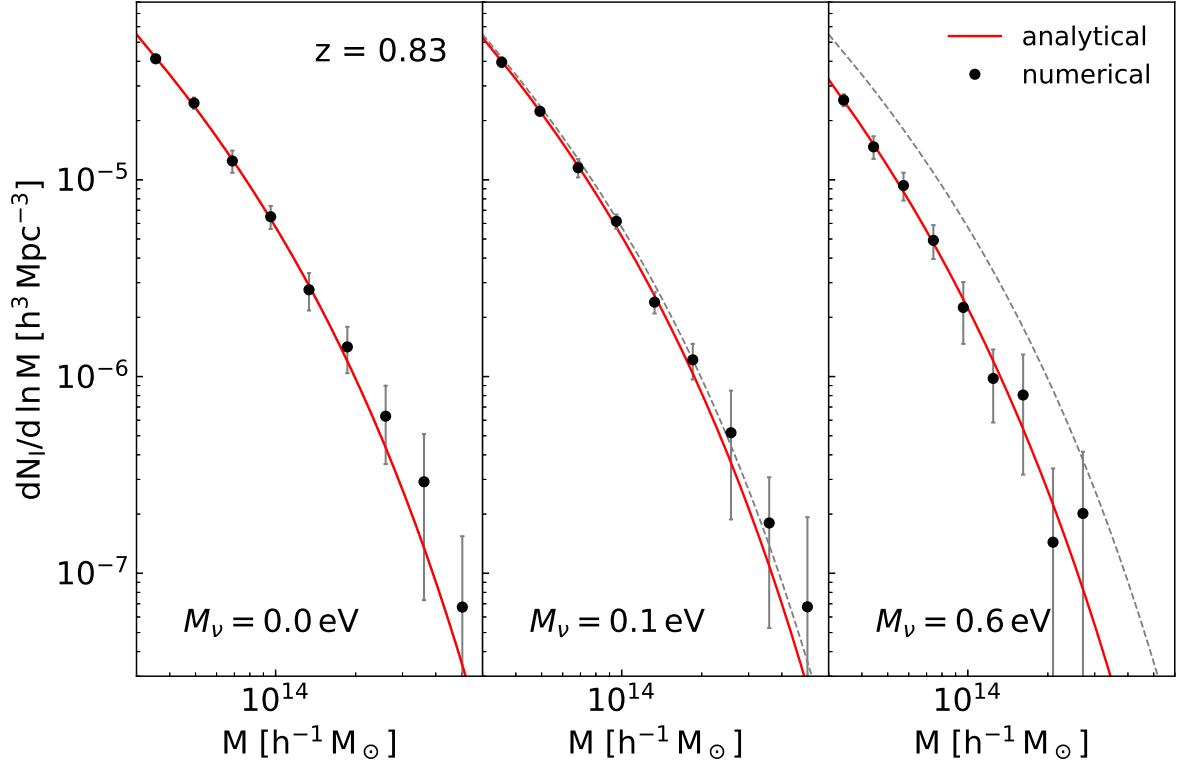}
\caption{Same as Figure \ref{fig:mf_z0} but for at $z=0.83$.}
\label{fig:mf_z0.8}
\end{center}
\end{figure}

Figure \ref{fig:mf_z0} displays both of the numerical field cluster mass functions from the \texttt{MassiveNuS} (filled circles) and 
the analytic model with the best-fit value of $\beta$ (red solid lines) at $z=0$ for the three different cases of $M_{\nu}$.  The errors in the 
numerical determination of $dN_{\rm I}/d\ln M$ is calculated as one standard deviation from the mean averaged over eight Jackknife resamples, 
\citep{RL20}. The black dotted lines in the middle and right panels conform to the red solid line in the left panel. 
Figures \ref{fig:mf_z0.4}-\ref{fig:mf_z0.8} show the same as Figure \ref{fig:mf_z0} but at $z=0.42$ and $0.83$, respectively.  As can be seen, the 
analytical single parameter model for $dN_{\rm I}/d\ln M$ agrees excellently well with the numerical results at all redshifts for all of the three cases of 
$M_{\nu}$, confirming its validity even in the presence of massive neutrinos and proving its robustness as a physical model. 

Figures \ref{fig:mf_z0}-\ref{fig:mf_z0.8} clearly show that $dN_{\rm I}/d\ln M$ has a significantly lower amplitude for the case of 
$M_{\nu}=0.6\,{\rm eV}$ than for the other two cases of $M_{\nu}=0.0\,{\rm eV}$ and $M_{\nu}=0.1\,{\rm eV}$, between which almost 
no difference is found in $dN_{\rm I}/d\ln M$, no matter at what redshifts they are compared with each other. 
Although the difference in $dN_{\rm }/d\ln M$ between the two cases of $M_{\nu}=0.0\,{\rm eV}$ and $M_{\nu}=0.1\,{\rm eV}$ tends to 
slightly increase with $z$, the larger errors in the measurement of $dN_{\rm I}/d\ln M$ at higher redshifts weigh down their statistical significances. 
The comparison of Figures \ref{fig:mf_z0}-\ref{fig:mf_z0.8} with Figure \ref{fig:pk} indicates that the $M_{\nu}$-dependence of the field cluster mass 
function is almost entirely dictated by the $M_{\nu}$-dependence of $P(k,z)$. 
As mentioned in Section \ref{sec:review}, the cosmology-dependence of the field cluster abundance (including its $M_{\nu}$-dependence) has two 
different sources, $P(k,z)$ and $\beta$. The results shown in Figures \ref{fig:mf_z0}-\ref{fig:mf_z0.8}, however, imply that the former overwhelms the 
latter in shaping the $M_{\nu}$-dependence of the field cluster mass function, which in turn warns that the field cluster mass function would fail 
not only in constraining $M_{\nu}$ below the Planck constraint but also in breaking the $\sigma_{8}$-$M_{\nu}$ degeneracy. 

\begin{figure}
\begin{center}
\includegraphics[scale=0.7]{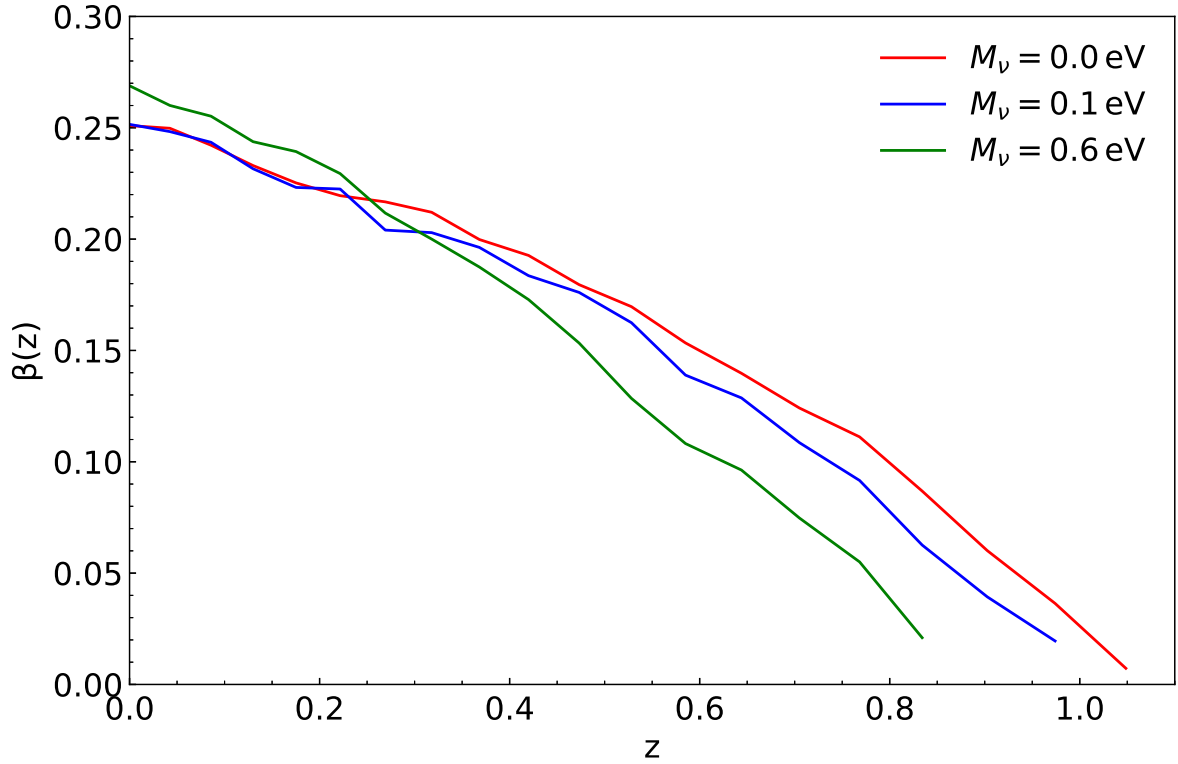}
\caption{Numerical results of the drifting coefficient, $\beta(z)$, in the redshift range of $0\le z \le 1$ for the 
three different cases of $M_{\nu}$, from the \texttt{MassiveNuS}.}
\label{fig:beta_z}
\end{center}
\end{figure}

Figure \ref{fig:beta_z} plots the numerically determined values of $\beta(z)$ at twenty different redshifts in the range of $0\le z\le 1$ for 
the three different cases of $M_{\nu}$, revealing that  $\beta(z)$ evolves differently among the three cases. Here, the errors, $\sigma_{\beta}$, are 
obtained through the Fisher information analysis, as done in \citet{RL20}. As can be seen, at $z\lesssim0.3$ the drifting coefficient $\beta(z)$ has 
higher values for the case of $M_{\nu}=0.6\,{\rm eV}$ than for the other two cases. Whereas at $z\gtrsim0.3$, the tendency is reversed. 
The most massive neutrinos case yields the lowest values of $\beta(z)$, while its highest values are found for the massless neutrinos case.  
In addition, we find that the slope of $\beta(z)$ substantially differs even between the two cases of $M_{\nu}=0.0\,{\rm eV}$ and 
$M_{\nu}=0.1\,{\rm eV}$, while no difference found in $\beta(z=0)$ between them. 

We speculate that this redshift-dependence of the effect of massive neutrinos on $\beta(z)$ might help break the $\sigma_{8}$-$M_{\nu}$ degeneracy. 
Recall that the effect of massive neutrinos on the linear density power spectra and cluster mass function is consistent in its direction, 
regardless of the redshifts, as witnessed in Figures \ref{fig:pk}-\ref{fig:mf_z0.8}. The more massive neutrinos always reduce more severely the amplitudes 
of $P(k,z)$ and $dN_{\rm I}/d\ln M$ at all redshifts, which is why the two diagnostics suffer from the $\sigma_{8}$-$M_{\nu}$ degeneracy. In other words, 
the lower value of $\sigma_{8}$ has the same effect on $P(k,z)$ (and $dN/d\ln M$ as well) as the higher value of $M_{\nu}$. 
Meanwhile, our result shown in Figure \ref{fig:beta_z} implies that the effect of the higher value of $M_{\nu}$ on $\beta(z)$ might be differentiated 
from that of the lower value of $\sigma_{8}$ on $\beta(z)$.  The latter lowers the amplitude of $\beta(z)$ without changing its slope, while the former 
heightens  its amplitude and concurrently steepens its slope.  Yet, the possibility of breaking the $\sigma_{8}$-$\Omega_{m}$ degeneracy with $\beta(z)$ is 
only a speculation, since we have yet to demonstrate its feasibility in practice.

\begin{figure}
\begin{center}
\includegraphics[scale=0.7]{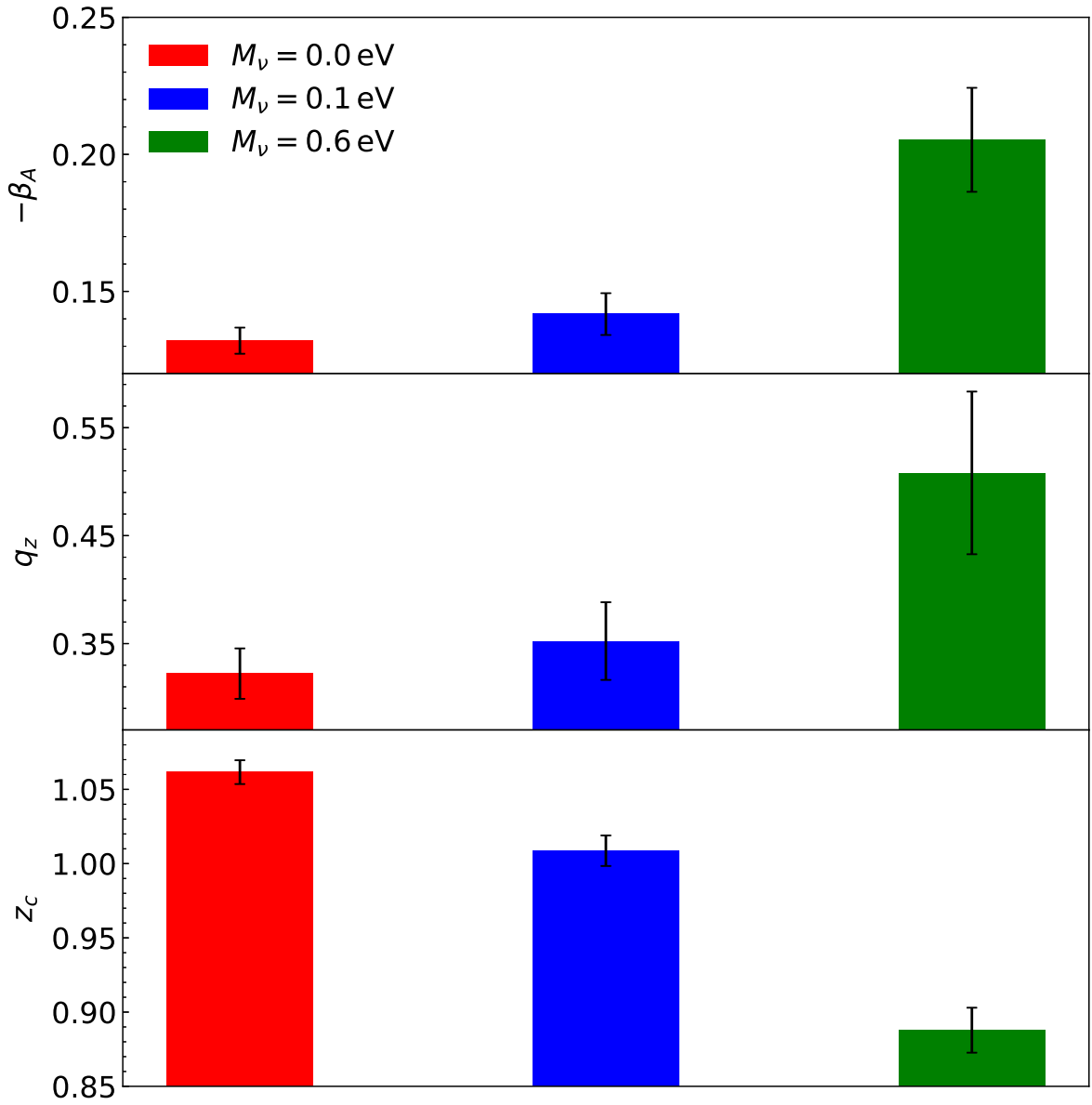}
\caption{Best-fit three parameters of the analytic formula for $\beta(z)$ for the three different cases of $M_{\nu}$.}
\label{fig:beta_sig}
\end{center}
\end{figure}

As done in \citet{RL20}, to effectively quantify the differences in the evolution of the drifting coefficient among the three cosmologies, 
we fit Equation (\ref{eqn:beta_fit}) to the numerically determined $\beta(z)$ by adjusting the values of $\beta_{A},\ q_{z}$ and $z_{c}$ 
to yield the minimum $\chi^{2}$. 
Figure \ref{fig:beta_fit} demonstrates how well the simple formula (red solid lines), Equation (\ref{eqn:beta_fit}), suggested by \citet{RL20}, 
agrees with the numerically obtained $\beta(z)$ (black filled circles) for all of the three cases of $M_{\nu}$. 
Figure \ref{fig:beta_sig} shows the best-fit values of $-\beta_{A},\ q_{z},\ z_{c}$ with their errors $\sigma_{\beta_{A}},\ \sigma_{qz},\  
\sigma_{zc}$, which are all obtained through the $\chi^{2}$ fitting after due consideration of the uncertainties in $\beta(z)$ shown 
in Figure \ref{fig:beta_fit}. 

The most significant differences among the three cases are found in the values of $z_{c}$, which is consistent with the result of \citet{RL20} 
that $z_{c}$ was found to vary most sensitively with the dark energy equation of state.  
Assessing the statistical significances of the differences in $z_{c}$ among the three cases of $M_{\nu}$ 
by estimating the errors of their mutual differences, $\sigma_{\Delta(zc)}$, propagated from  $\sigma_{zc}$, as done in \citet{RL20}, 
we find the difference in $z_{c}$ between the two cases of $M_{\nu}=0.0\,{\rm eV}$ and $M_{\nu}=0.1\,{\rm eV}$ ($M_{\nu}=0.6\,{\rm eV}$) 
to exceed $4\sigma_{\Delta({zc})}$  ($10\sigma_{\Delta (zc)}$). Whereas, the differences in the other two parameters, $\beta_{A}$ and $q_{z}$, 
between the two cases of $M_{\nu}=0.0\,{\rm eV}$ and $M_{\nu}=0.1\,{\rm eV}$ ($M_{\nu}=0.6\,{\rm eV}$) are found to be statistically insignificant 
(not so significant as that in $z_{c}$). 

It should be worth explaining here why $z_{c}$ is the most sensitive to the variation of $M_{\nu}$.   Given the definition $z_{c}$ as a critical redshift at which 
$\delta_{c}=1.686$ (i.e, $\beta(z_{c})=0$), its value should be determined by two factors, both of which sensitively depend on $M_{\nu}$. 
The first factor is how fast the matter density parameter $\Omega_{m}$ approaches unity (i.e, the Einstein-de Sitter value) at high redshifts, while
the second one is how rare the field clusters are in a given universe, since the gravitational collapse of the rarer objects proceeds in a more spherically 
symmetrical way \citep{ber94}.  Meanwhile, the other two parameters, $\beta_{A}$ and $q_{z}$, depend mainly on either of the two factors: $\beta_{A}$ on the 
second, while $q_{z}$ on the first.  

\section{Discussion and Conclusion}\label{sec:con}

Conducting a numerical analysis of the \texttt{MassiveNuS} data \citep{massivenu}, we have found that the massive neutrinos have a unique redshift-dependent 
effect on the drifting coefficient of the field cluster mass function, $\beta(z)$, which measures the difference between the density thresholds for the realistic 
nonspherical and the idealistic EdS spherical collapse at a given mass scale.  In our previous work \citep{RL20}, we already found that $\beta$ vanishes to zero 
at a certain critical redshift $z_{c}$ but increases as the universe evolves from $z_{c}$ to $z=0$ like an inverse sine hyperbolic function of $z$. 
We have newly found here that the presence of more massive neutrinos lowers $z_{c}$ and induce a faster increase of $\beta(z)$ with the decrement 
of $z$ below $z_{c}$. The $\nu\Lambda$CDM cosmology with total neutrino mass of $M_{\nu}=0.6\,{\rm eV}$ has been found to yield higher (lower) values 
of $\beta$ at $0\le z\lesssim z_{\rm th}$ ($z_{\rm th}\lesssim z\le z_{c}$) than the $\Lambda$CDM cosmology with massless neutrinos with $z_{\rm th}\sim 0.3$.  
Noting that this redshift-dependent effect of massive neutrinos on $\beta$ is quite unique and distinct especially from the redshift-independent effect of 
$\sigma_{8}$ on $\beta(z)$,  we suggest that the drifting coefficient of the field cluster mass function should allow us to break the notorious $\sigma_{8}$-$M_{\nu}$ 
degeneracy, which has haunted for long the conventional probes of $M_{\nu}$ based on the linear density power spectrum. 

Our physical explanation for this distinct redshift-dependent effect of $M_{\nu}$ on $\beta(z)$ is that it is generated by a competition between 
the suppressed small-scale powers and the increased degree of the anisotropy of the cosmic web in the presence of massive neutrinos. 
As shown by \citet{ber94}, the formation of a rare event like a massive cluster (or a field cluster) is well approximated by a spherical collapse 
process. The rarer an object is, the more spherically its gravitational collapse proceeds. 
In the presence of more massive neutrinos which suppress more severely the small-scale powers, a field cluster corresponds to an even rarer object 
since it originates from a more extreme local maximum in the initial density field.   Therefore, it is naturally expected that in the presence of 
more massive neutrinos the collapse density threshold $\delta_{c}$ for the field clusters would become closer to the spherical threshold 
$\delta_{sc}$ (or equivalently, $\beta$ closer to zero). 

The free streaming of massive neutrinos, however, has another effect of rendering the cosmic web more anisotropic in the deeply nonlinear stage. 
According to the previous works \citep[e.g.,][]{shi-etal14,ho-etal18} which found the degree of the anisotropy of the cosmic web to depend on the background 
cosmology, the stronger gravity at a given scale pulls down the anisotropic feature of the cosmic web in the nonlinear stage. 
The free streaming of massive neutrinos plays a role along with DE in weakening the gravitational clustering on the cluster scale,  which in consequence 
increases the degree of the anisotropy of the cosmic web.  The stronger tidal influences from the more anisotropic cosmic web \citep{cosmicweb} 
deviate the collapse process further from the spherical symmetry, elevating $\beta$ above zero. 

\begin{figure}
\begin{center}
\includegraphics[scale=0.7]{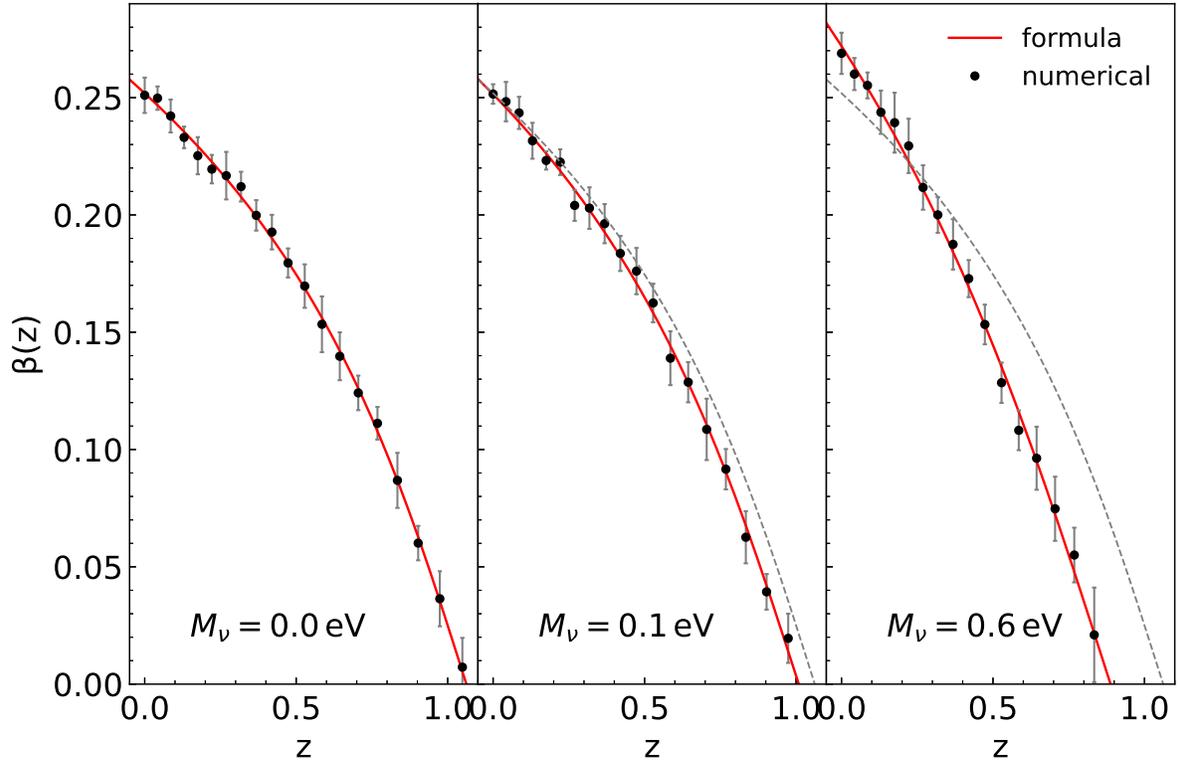}
\caption{Best-fit formula for $\beta(z)$ (red solid line) over-plotted with the numerical results (filled circles) 
for the three different cases of $M_{\nu}$. The dotted lines in the middle and right panels conform to the 
red solid line in the left panel.}
\label{fig:beta_fit}
\end{center}
\end{figure}

At high redshifts ($z_{\rm th} \lesssim z \le z_{c}$), the first effect of massive neutrinos overwhelms the second, lowering $\beta$ close to zero,  
since the high-$z$ field clusters correspond to the rarest events formed through the collapses of the highest density peaks which proceed in almost 
perfectly spherically. However, at lower redshifts ($0\le z\lesssim z_{\rm th}$) after the onset of the nonlinear evolution of the cosmic web, 
the second effect wins over the first, deviating $\beta$ further from zero.   Our result shown in Figure \ref{fig:beta_fit} reveals the $M_{\nu}$-dependence 
of the threshold redshift, $z_{\rm th}$, at which the second effect becomes more dominant than the first. It is around $0.3$ for the case of 
$M_{\nu}=0.6\,{\rm eV}$, while it becomes around zero for the case of $M_{\nu}=0.1\,{\rm eV}$.  The more massive neutrinos induce the turn-over of 
the second effect to occur earlier.  Our future work is in the direction of constructing a more theoretical model for $\beta(z)$, within which the 
$M_{\nu}$-dependences of $z_{\rm th}$ and $z_{c}$ could be predicted.

Another important hint of this work is that the sensitivity of $\beta(z)$ to $M_{\nu}$ might be high enough to detect the effect of massive neutrinos on it, 
even in case that $M_{\nu}$ is as low as $0.1\,{\rm eV}$ below the Planck constraint \citep{planck18}.  
The signal of the difference in $z_{c}$ between the $\Lambda$CDM and $\nu\Lambda$CDM with $M_{\nu}=0.1\,{\rm eV}$ 
($M_{\nu}=0.6\,{\rm eV}$) cosmologies has been found to be approximately four (ten) times higher than the propagated errors. 
Given that the observational data from much larger volumes than that of the \texttt{MassiveNuS} are already in the pipeline 
\citep[e.g.,][]{euclid19},  we conclude that  the drifting coefficient of the field cluster mass function, $\beta(z)$, 
has a good prospect for providing a very powerful complementary probe of $M_{\nu}$ in practice. 

\acknowledgments

We are grateful to an anonymous referee for useful comments. 
We thank the Columbia Lensing group (http://columbialensing.org) for making their suite of simulated maps available, and NSF for supporting the creation 
of those maps through grant AST-1210877 and XSEDE allocation AST-140041. We thank New Mexico State University (USA) and Instituto de Astrofisica de 
Andalucia CSIC (Spain) for hosting the Skies \& Universes site for cosmological simulation products.

We acknowledge the support by Basic Science Research Program through the National Research Foundation (NRF) of Korea 
funded by the Ministry of Education (No.2019R1A2C1083855) and also by a research grant from the NRF to the Center for Galaxy 
Evolution Research (No.2017R1A5A1070354).

\clearpage
\clearpage
\clearpage
\clearpage
\clearpage
\end{document}